\documentclass[12pt,a4,pdftex,revtex,amsfonts,amsmath,amssymb]{article}

\usepackage{color}
\usepackage{soul}
\usepackage{cite}
\usepackage{amsmath}
\usepackage[dvips]{graphics}
\usepackage{graphicx}
\usepackage{multicol}
\usepackage{cleveref}
\usepackage[makeroom]{cancel}

\definecolor{darkgreen}{rgb}{0,0.5,0}

\usepackage{cancel}

\advance \topmargin by -\headheight
\advance \topmargin by -\headsep
\advance \topmargin by -1.0cm
\evensidemargin \oddsidemargin
\evensidemargin \oddsidemargin
\advance\hoffset by -3mm  
\advance\voffset by  8mm  
\newlength{\figwidth}
\newlength{\figheigth}
\newcommand*{\revadd}[1]{#1}
\newcommand*{\revrem}[1]{}
\newcommand*{\comment}[1]{}

\newcommand{\be}{\begin{equation}}
\newcommand{\ee}{\end{equation}}
\newcommand{\bea}{\begin{eqnarray}}
\newcommand{\eea}{\end{eqnarray}}
\newcommand{\nn}{\nonumber}

\newcommand{\rr}{\mathbf{r}}
\newcommand{\RHO}{\mathbf{\rho}}

\newcommand{\rhorom}{{\rho\left({\mathbf r},\Omega\right)}}

\newcommand{\F}{{\cal F}}

\newcommand{\PP}{\mathbf{P}}

\begin{document}

\begin{center}
{\Large Fast Computation of  Solvation Free Energies  
with Molecular Density Functional Theory:
Thermodynamic-Ensemble  Partial Molar Volume Corrections}

\vspace{5mm}

{\bf  Volodymyr P. Sergiievskyi, Guillaume Jeanmairet, Maximilien Levesque, \\ and Daniel Borgis}

\vspace{5mm}

{\em P\^ole de Physico-Chimie Th\'eorique, \'Ecole Normale Sup\'erieure, UMR 8640 CNRS-ENS-UPMC, 24, rue Lhomond, 75005 Paris, France}

\end{center}



\begin{abstract}
Molecular Density Functional Theory (MDFT) offers an efficient implicit-solvent method to estimate molecule solvation free-energies 
whereas  conserving a fully molecular representation of the solvent. Even within a second order approximation for the free-energy functional, the so-called homogeneous reference fluid approximation, we show that the hydration free-energies computed for a dataset of 500 organic compounds   are of similar quality
as those obtained from  molecular dynamics free-energy perturbation simulations, with a computer cost reduced by two to three orders of magnitude. 
This requires to introduce the proper partial volume correction to transform the results from the grand canonical to the isobaric-isotherm ensemble that is pertinent to  experiments.  We show that this correction can be extended to 3D-RISM calculations, giving a sound theoretical justification to empirical partial molar volume corrections that have been proposed recently.
\end{abstract}

\newpage

\section{Introduction}

Solvation Free Energy (SFE) is one of the main physical quantities in solution chemistry. 
Many important characteristics, 
such as dissociation constants, partition coefficient (log P),  
which are necessary for describing most of the processes in 
physical chemistry and biochemistry are expressed through the SFE.
Despite the importance of that physical quantity, determination of SFE is often problematic.
Experimental determination of SFE is often complicated.
It can require essential time and resources, especially if SFE is calculated for low soluble and low volatile substances\cite{perlovich_solvation_2004,perlovich_towards_2006}.
This increases the importance of the numerical SFE calculations.
SFE calculation methods can be separated into two classes: (i) explicit solvent methods (simulations)\cite{frenkel_understanding_2002,matubayasi_free-energy_2009}, and (ii) implicit solvent methods \cite{tomasi_quantum_2005}.
As for the advantages of the simulation methods we can name their relatively high accuracy (however, one should remember that accuracy of the simulations greatly depend on the force-field and partial charges determination) \cite{david_l_mobley_small_2009,matubayasi_free-energy_2009,konig_predicting_2012}.
One of the disadvantages of the explicit solvent methods is their high demands to the computational resources, which make them inapplicable in some practical applications where the speed is critical.

Among the most common implicit solvent methods are the continuum electrostatics models, which are based on solutions of the Poisson-Boltzmann equation for the charges inside the molecular cavity inside the dielectric continuum \cite{tomasi_quantum_2005,warwicker_calculation_1982}.
This type of methods allows one to perform the calculations much faster than it is done in the simulations. 
However, accuracy of implicit solvent methods is often not enough for accurate prediction of the SFE; 
that is why in practice the methods with empirical corrections are often used. 
Many of methods of this type, such as COSMO-RS or SM6/SM8, allow one to calculate SFE with a high accuracy for some classes of simple compounds \cite{cramer_universal_2008,labute_generalized_2008,nicholls_predicting_2008,phillips_quantum_2008,klamt_prediction_2003}.
However, for the compounds with complicated structure these methods often fail to give a good correspondence to experiments \cite{klamt_prediction_2009,marenich_performance_2009}.

On the other hand, the numerical  methods that have emerged in the second part of the last century from  liquid-state theories\cite{hansen_theory_2000, gray-gubbins-vol1}, including integral equation theory in the reference interaction-site model (RISM) approximation\cite{chandler_optimized_1972,hirata-rossky81,hirata-pettitt-rossky82,reddy03,pettitt07,pettitt08}
or in the molecular picture\cite{fries-patey85,richardi98,richardi99}, classical density functional theory (DFT)\cite{evans79,evans92,hansen_theory_2000,Wu07}, or classical fields theory\cite{chandler93,lum99}, become methods of choice for 
many physical chemistry or chemical engineering applications\cite{gray-gubbins-vol2,neimark06,neimark11,wu06}. They can be used as evolved implicit solvent methods to predict the solvation properties of molecules  at a much more modest computational cost than  molecular dynamics (MD) or Monte-Carlo (MC) simulations, whereas retaining the molecular  character of the solvent. 
 There have been a number of recent efforts in that direction using 3D-RISM\cite{Beglov-Roux97,kovalenko-hirata98,red-book,yoshida09,kloss08-jcp,kloss08-jpcb}, lattice field theories\cite{azuara06,azuara08} or Gaussian field theories\cite{lum99,tenwolde01,huang02,varilly11}. Another important class of approaches relies on  classical  DFT in the molecular\cite{ramirez02,ramirez05,gendre09,zhao11,borgis12,levesque12_1,levesque12_2,jeanmairet13,jeanmairet13_2,zhao-wu11,zhao-wu11-correction} or interaction site\cite{liu13} representation, or constructed from the  statistical associating fluid theory  (SAFT)\cite{roundy13}.
 
3D-RISM in the hyper-netted chain (HNC) or Kovalenko-Hirata (KH) approximation is certainly the method that has been pushed the furthest in that direction to date, with recent application to the high-throughout prediction of organic molecules solvation free-energies\cite{ratkova_accurate_2010,palmer_towards_2010,frolov_hydration_2011,truchon_cavity_2014}, and the development of highly efficient multigrid algorithms for that purpose\cite{sergiievskyi_multigrid_2011,sergiievskyi_3DRISM_2012,luchko_three-dimensional_2010}. Unfortunately,  it turns out that SFE's calculated that way systematically overestimate the experimental values, and display a very poor correlation
\cite{lue_liquid-state_1992,chuev_improved_2007}.
It was found that the accuracy can be considerably improved by including  an empirical partial molar volume (PMV) correction \cite{chuev_improved_2007,palmer_towards_2010,frolov_hydration_2011,ratkova_accurate_2010}, {\em i.e.} a term of the form $a \Delta V + b$, where $a$ and $b$ are adjustable parameters. Truchon et al. have proposed very recently a rationalization of these PMV corrections using a physically-motivated form of the factor $a$, but with yet an adjustable multiplicative parameter.\cite{truchon_cavity_2014}
 
Towards similar goals,   a molecular density functional theory (MDFT) approach to solvation has been
introduced recently ~\cite{ramirez02,ramirez05,gendre09,zhao11,borgis12,levesque12_1,levesque12_2,jeanmairet13,jeanmairet13_2}.
 It relies on the definition of a free-energy functional depending on the full six-dimensional position and orientation solvent density. In the so-called homogeneous reference fluid (HRF) approximation, the (unknown) excess free energy can be inferred from the angular-dependent direct correlation function of the bulk solvent,
that can be predetermined from molecular simulations of the pure solvent.\cite{ramirez05-CP,zhao-borgis13} This is equivalent to a second order Taylor expansion of the excess free-energy around the homogeneous liquid density. In a recent work\cite{jeanmairet13}, we  introduced an even simplified version of MDFT for water, that can be derived rigorously for simple point charge representations of water such as SPC or TIP4P, involving a  single Lennard-Jones interaction site and distributed partial charges. In that case we showed that the functional can be expressed in terms of the particle density $n(\rr)$ and site-distributed polarisation density $\PP(\rr)$, and it requires  as input three simple bulk physical properties of water, namely the density structure factor, and the k-dependent longitudinal and transverse dielectric susceptibilities. Those quantities can be inferred from experiments, or from molecular dynamics simulations of the selected point-charge model in bulk conditions\cite{bopp96,bopp98}. 

In Refs.~\cite{zhao11,borgis12,levesque12_2,jeanmairet13,jeanmairet13_2}, MDFT was applied to the solvation structure and solvation thermodynamics of molecular solutes in acetonitrile and water. For water, it was shown that the inclusion of three-body corrections beyond the straight HRF approximation was necessary to describe accurately  the hydration free-energy of hydrophobes\cite{levesque12_2,jeanmairet13_2}, or the water structure around chemical groups giving rise to strong hydrogen bonding to the solvent\cite{jeanmairet13}. In the same context, Wu and collaborators have applied recently their interaction-site DFT approach to the  high-throughout prediction of the solvation free-energies of neutral organic molecules.\cite{liu_high-throughput_2013} They showed that excellent correlation to experimental or MD results can be reached using an isotropic hard-sphere bridge corrections (as in Refs.~\cite{zhao-wu11,zhao-wu11-correction,liu13} from the same group; see also 
Refs.~\cite{levesque12_2,jeanmairet13_2}). Here, keeping in mind all possible refinements,  we go back to the straight HRF approximation for the same problem, and rather focus on the application of our  theory to the computation of SFE's in the proper thermodynamic ensemble corresponding to the experimental conditions. Doing so, we are able to  draw a link between our theoretical approach and the empirical PMV corrections proposed in 3D-RISM calculations, since the HRF approximation in classical DFT is equivalent to the HNC approximation in integral equation theories.

\section{Sketch of the theory}



In  molecular density functional theory, the solvent molecules  are considered as rigid entities with position $\rr$ and orientation $\Omega$ (in terms of three Euler angles $\theta, \phi, \psi$) and are described by an atomistic force field. The solute, described within the same atomistic force field as a collection of Lennard-Jones sites carrying partial charges,  creates at each point in the solvent an external microscopic potential $v(\rr,\Omega)$.  The system is then characterized by the position and orientation-dependent inhomogeneous solvent density $\rhorom$ and by a free-energy functional $\F[\RHO] \equiv \F[\rho(\rr,\Omega)]$, which can be written  as  \cite{hansen_theory_2000}
\begin{equation}
\label{eq:F}
  \mathcal{F}[\RHO] =
  kT \int \left[ \rho(1) \ln \left(  \Lambda_{rot}  \Lambda^3 \rho(1)  \right) - \rho(1) \right] d1  
  + \int \rho(1) \, v(1) d1
  + \mathcal{F}^{exc}[\RHO].
\end{equation}

In short notations,  $(1)$ stands for $(\rr_1,\Omega_1)$, and $d1 = d\rr_1d\Omega_1$.
The first term is the ideal free-energy, with $\Lambda$ the de Broglie thermal wave length; with respect to standard expression for isotropic particles\cite{evans79,evans92,hansen_theory_2000}, the additional $\Lambda_{rot}$ term corresponds to the inclusion of the orientational degrees of freedom
\footnote{
\revadd{
$\Lambda_{rot} = (2\pi \hbar^2)^{3/2} (J_{x} J_{y} J_{z})^{-1/2} kT^{-3/2}$
where $J_{x}$, $J_y$, $J_z$ are  moments of inertia of the molecule about the principal axes. See Ref. \cite{volodymyr_sergiievskyi_modelling_2013}(section 3.2) for the derivation. 
}
}.
 
$\F^{exc}[\RHO]$ is the excess free energy functional, which accounts for particle effective interactions in the liquid. In the homogeneous reference fluid approximation (HRF)\cite{ramirez02,ramirez05}, reminiscent of the hyper-netted chain approximation (HNC) in integral equation theories,  the excess free energy functional is represented by the first two terms of the Taylor series expansion around the homogeneous liquid state at the  density $\rho_0=n_0/8\pi^2$ ($n_0$ the particle number density), {\em i.e.},
\begin{equation}
\label{eq:F_exc}
  \mathcal{F}^{exc} [\RHO]
  =
  \mathcal{F}^{exc}[\rho_0]
  +
  \int 
      \left.
      \frac{\delta \mathcal{F}^{exc}}
           {\delta \rho(1)}
      \right|_{\rho_0}     
           \Delta \rho(1) d1
  +
  {1 \over 2}
  \int \Delta \rho(1) 
       \left.
       \frac{ \delta^2 \mathcal{F}^{exc}}
            {\delta \rho(1) \delta \rho(2) }
       \right|_{\rho_0}     
            \Delta \rho(2) 
            d1d2,
\end{equation}
where
$\Delta \rho(1) \equiv \rho(1) - \rho_0$,
and by definition,  
$
 	   \left.
       {\delta^2 \mathcal{F}^{exc}} /
       {\delta \rho(1) \delta \rho(2) }
       \right|_{\rho_0}
       \equiv
       -kT c(12),
$ 
where $c(12)$ is the angular dependent  two-particle direct correlation function of the pure solvent
\footnote{
\revadd{
In fact, HNC approach is just a formulation of the HRF approximation in terms of integral equations, and vice versa: HRF is a formulation of the HNC theory in terms of DFT.
This can be proven by the fact that functional derivative of \eqref{eq:F} w.r.t $\rho(1)$ just gives the HNC closure. 
}
}
. 

The correct thermodynamic
ensemble to be considered in  classical DFT is the grand-canonical ensemble in which  the chemical potential of the solvent,
$\mu_0$, is imposed.  One thus has to deal with the grand-potential
\be
\label{eq:Theta}
\Theta[\RHO] = \F[\RHO] - \mu_0 \int  \rho(1) \, d1.
\ee
The value of the chemical potential is imposed, for example, by direct contact with a huge reservoir at constant density $\rho_0$.
The relation between $\mu_0$ and $\rho_0$ is obtained from the requirement that, in the absence of external perturbation, the density in the system should be the uniform density $\rho_0$, 
{\em i.e.}, $\left. \revadd{\delta} \Theta/\delta \rho(1)\right|_{\rho_0} = 0$, yielding
\be
\label{eq:mu_0}
\mu_0 = \mu_0^{id} + \mu_0^{exc} =
 kT \ln \Lambda_{rot} \Lambda^3 \rho_0 + 
 \left.
      \frac{\delta \mathcal{F}^{exc}}
           {\delta \rho(1)}
 \right|_{\rho_0}.
\ee
\revadd{We note, that $\delta \mathcal{F}^{exc} / \delta \rho(1). $ is independent on the particle coordinates due to homogeneity of the fluid in absence of the external potential.}

For a given solute creating a potential $v(1)$,  the grand-canonical solvation free-energy is obtained  by minimizing 
the functional $\Delta \Theta[\RHO] = \Theta[\RHO] - \Theta[\RHO_0]$, which, according to \cref{eq:F,eq:F_exc,eq:Theta,eq:mu_0}  
amounts to the following expression that we used in all of our previous works; see supporting information for derivation.
\bea
\Delta \Theta[\RHO] &=&  kT \int \left(
     \rho(1) \ln ( {\rho(1) \over \rho_0} ) 
     - 
     \Delta \rho(1) 
     \right) d1
 + 
 \int \rho(1) v(1) d1 \nn \\
 & & -
  {kT \over 2}
  \int \Delta \rho(1) c(12)
            \Delta \rho(2) 
            d1d2.         
\eea
Such functional form insures that $\rhorom = \rho_0$ far from the solute where $v(\rr,\Omega)=0$. Its minimization yields the inhomogeneous equilibrium density of the solvent in the presence of the solute and the solute grand-canonical solvation free-energy, $\Delta \Theta_{MDFT}$. The latter is evaluated with respect to the homogeneous fluid at the same chemical potential $\mu_0$ (thus staying all the way in contact with the reservoir at constant density $\rho_0$) and at the same constant volume $V$. This corresponds to the transition between state 1 and state 2 in Figure 1.
In the initial state, the system contains $N + \Delta N = \int_V  n_0 d\rr =  n_0 V$, whereas in the final one it contains $N = \int_V d\rr \, n(\rr)$, where the number density is defined by $n(\rr) = \int d\Omega \, \rhorom$.

\begin{figure}[h]
\begin{center}
  \includegraphics[width=0.8\textwidth]{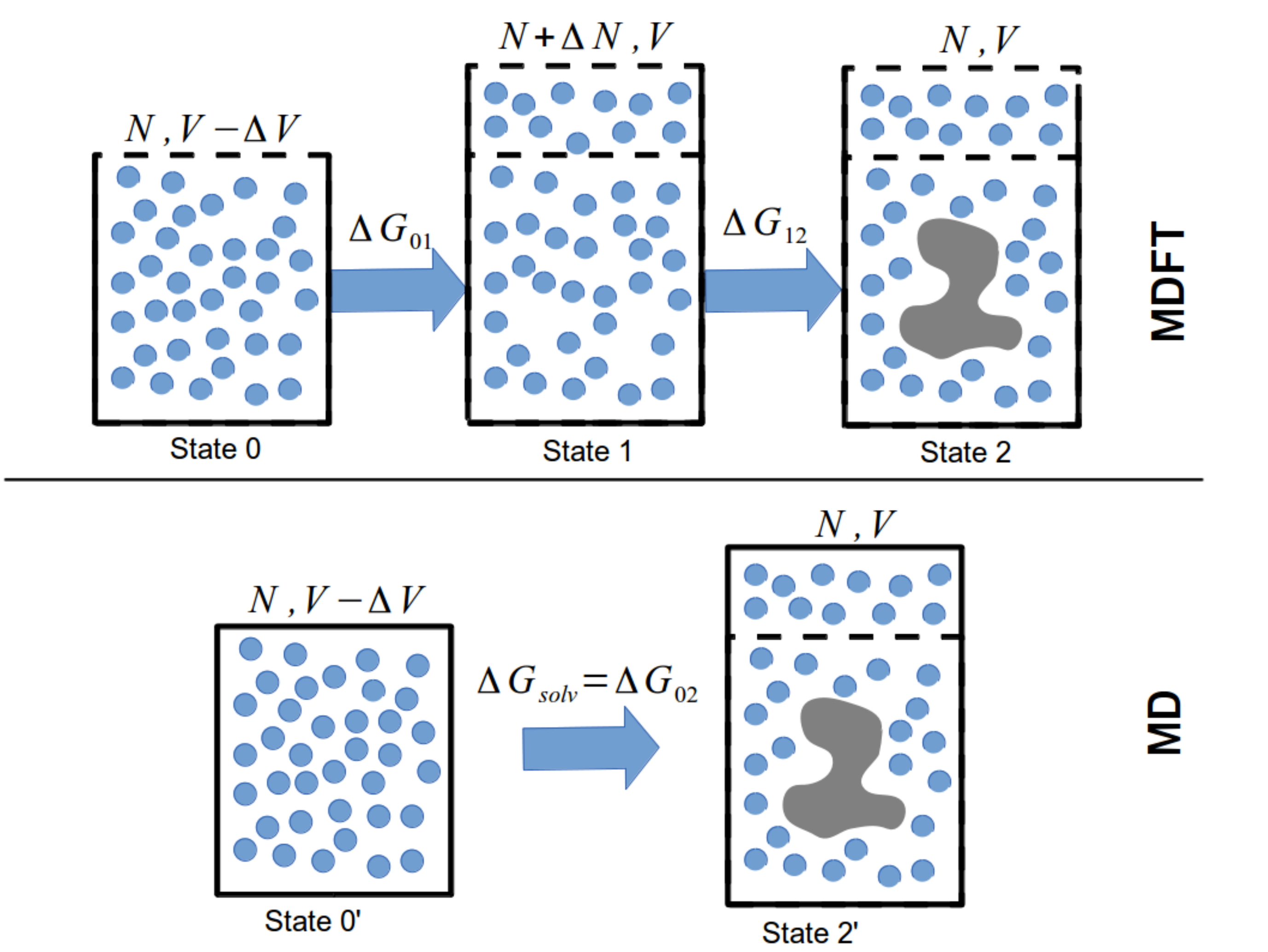}
\end{center}
   \caption{\label{fig:MDFT_vs_MD}
Thermodynamic scheme for the computation of the solvation free energy using the MDFT in the grand-canonical ensemble (top) or the molecular dynamics free energy perturbation method in the isobaric-isotherm ensemble (bottom). In the thermodynamic limit, states 0 and 0' and states 2 and 2' are equivalent, which motivates \cref{eq:Delta G fixed,eq:mu solute} in the text.
 }
\end{figure}

For a given solute model, this solvation free-energy should be compared to either the known experimental values, or to the results of 
 explicit molecular simulations,
 both generally determined in the isobaric-isotherm ensemble with a  constant number of solvent molecules, $N$, and a constant pressure, $P$. 
  To be consistent between  theoretical approaches, let us recall how  free energies can be computed in simulations.
 There are several methods based on the MD or MC simulations for calculation of the free energy differences between two given states.
 Commonly used are the free-energy perturbation  \cite{robert_w._zwanzig_high-temperature_1954},
 thermodynamic integration \cite{kirkwood_statistical_1935,frenkel_understanding_2002} and Bennett acceptance ratio 
 \revadd{
\cite{bennett_efficient_1976,shirts_solvation_2005,pham_identifying_2011}
 }
  methods.
The common feature of all these methods is that in addition to the simulations in the initial and final states one needs to perform the simulations in a series of intermediate states. 
In case of the  free-energy perturbation calculations,  the following standard scheme may be used, as in Ref.~\cite{david_l_mobley_small_2009}:
 A  cubic box is considered and the solute  placed at the center, the remaining space is filled with solvent particles, which amount to a fixed number $N$; the simulation box volume is further  released to accommodate for the imposed pressure. We denote as $V$ the volume of the box after the relaxation; this corresponds to state $2'$ in Fig.~1. The solute-solvent interactions  are then progressively turned off and the corresponding thermodynamic work is estimated. The system comes to  state 0' in Fig.~1, which corresponds to the uniform fluid in a volume $V - \Delta V$, with a homogeneous density   $n_0 = N/(V - \Delta V)$ corresponding to that of the chosen solvent model.  The free-energy of this transformation is just the inverse of the solute Gibbs SFE, $\Delta G_{solv}$, to be compared with the experimental value.

 It is clear then from this overall scheme  that the MDFT solvation process should be evaluated using the same reference state, the uniform fluid  of density 
 $n_0$ in a  volume $V - \Delta V$, state 0 in Fig.~1, that is identical to state 0' in the thermodynamic limit with negligible particle number and volume fluctuations. 
 To be compared with $\Delta G_{solv}$,  the free-energy difference computed by minimization, $\Delta G_{12}$, 
 should  be corrected by the free energy difference between state 1 and 0  in Fig.~1, $\Delta G_{01}$, involving a volume change from $V-\Delta V$ to $V$ at constant density, {\em i.e.}, 
 \be
 \label{eq:G_solv}
 \Delta G_{solv} = \Delta G_{02}  = 
 \Delta G_{01} + \Delta G_{12}
 \ee
This expression can be rewritten in terms of Helmholtz free energy
\be
\label{eq:G_solv2}
 \Delta G_{solv} = \Delta F_{01} + \Delta F_{12} - W_{01} - W_{12} = \Delta F_{01} + \Delta F_{12}
\ee
where $\Delta F_{01}$, $\Delta F_{12}$ are Helmholtz free energy changes and $W_{01}$, $W_{02}$ are the pressure mechanical works of the system in the  transitions 
$0 \to 1$ and $1 \to 2$, respectively. The second equality stems from the fact that the pressure works exactly compensate between the solvent expansion in $0 \to 1$ and its compression by the solute insertion in $1 \to 2$. See the supporting information for complete proof. As detailed there too,  injection of the zero density $\rho=0$ in \cref{eq:F_exc} with the condition that $\F_{exc}[\rho =0] = 0$, and use of  eqs (\ref{eq:F})-(\ref{eq:mu_0}) gives access to the value of the Helmholtz free energy in the volume $V$ at constant uniform, position and orientation density $\rho_0$ 
\be 
\label{eq:F0}
F_V[\rho_0]
   =  - P_0 V + \mu_0 N_0 = - \left(  n_0 kT - \frac{kT}{2}  n_0^2  \hat{c}_S(k=0)  \right) V 
    + \mu_0 N_0,
\ee 
equalities that defines  the homogeneous pressure $P_0$, and where $N_0 = n_0 V$.  $c_S(r_{12}) = \int d\Omega_1, d\Omega_2 c(\rr_{12}, \Omega_1, \Omega_2) $ represents the spherical component of the solvent direct correlation function,
and $\hat c_S(k=0) \equiv \int c_S(|\rr|) d\rr$. Obviously, the corresponding Gibbs free energy is given by  $G_V[\rho_0] = \mu_0 N_0$. 
The above equation  leads to
\be
  \Delta F_{01} = \mu_0 \Delta N - \Delta N kT + \dfrac{kT}{2} n_0^2 \hat c_S(k=0) \Delta V.
 \ee
Furthermore, going from the grand-canonical to canonical ensemble yields
 \be
  \Delta F_{12} = \Delta \Theta_{MDFT} + \mu_0 (-\Delta N),
\ee
and we get  finally using \cref{eq:G_solv,eq:G_solv2}
\be
\label{eq:Delta G fixed}
 \Delta G_{solv} = 
 \Delta \Theta_{MDFT} -n_0  kT \Delta V + {kT \over 2} n_0^2 \hat c(k=0) \Delta V
\ee
One can easily relate the volume  increase $\Delta V$ in $0 \rightarrow 1$
 to the particle number variation in $1 \rightarrow 2$, $\Delta N = N - N_0$, {\em i.e.},
\be
\label{eq:PMV}
\Delta V =  -\frac{1}{n_0} \Delta N = \frac{1}{n_0} \left(N_0 -  \int d\rr \, n(\rr_1) \right).
\ee
This defines $\Delta V$ as  the {\em partial molar volume} of the solute at infinite dilution in the considered solvent.

We note that expression \eqref{eq:Delta G fixed} is only valid for the case when the solute has a fixed position and the bulk density is kept unchanged in the process $0 \to 1 \to 2$.
However, in experiments, and in most simulations, the solutes 
 are not fixed and they can be considered as forming a homogeneous solution (at infinite dilution) with the solvent.
\revrem{This fact does not change the energy itself, but it does affect the chemical potential of water.}
\revadd{As the volume of the system changes, the standard state volume correction to the thermodynamic quantities should be applied \cite{b._lee_relation_1994}. }
It can be easily realized  that \revrem{in a system with a mobile solute,} \revadd{if} the whole volume is accessible to the solvent \revrem{, and}   its  ideal chemical potential  becomes
\be
\revadd{
  \mu_2^{id} = kT \ln \Lambda^3 {\Lambda_{rot} \over \Phi} { N \over V}
  \neq
   kT \ln \Lambda^3 {\Lambda_{rot}  \over \Phi } 
   { N \over V - \Delta V }
   =\mu_0^{id},
}   
\ee
whereas the excess chemical potential  is unaffected
\footnote{
\revadd{
$\Phi = \int_{\Omega} d\Omega$ appeared in the eqiation because of different normalizations of the six-dimensional density $\rho(\rr,{\Omega})$ and number density $n(\rr) = \int \rho(\rr,\Omega) d \Omega$
}
}. 
Along this lines, since the Gibbs free energy in  state 0 is $G_0 = \mu_0 N$, the Gibbs free energy in  state 2 can be written as 
\bea
\label{eq:G2}
G_2  &=& \Delta G_{solv}  + \mu_0 N \nn \\
 &=& \Delta G_{solv} + \mu_2 N + \left( \mu_0 - \mu_2 \right) N \nn \\
 & = &  \Delta G_{solv} + \mu_2 N + \revadd{n_0 \Delta V kT}.
 \eea
We have used the fact that 
\revadd{
$\mu_0^{id} - \mu_2^{id} 
= NkT \ln ( V / (V-\Delta V) )
\approx NkT \Delta V / (V-\Delta V) = n_0 \Delta V kT 
$. The relation is strict when $V \to \infty$}.
On the other hand, the system with a \revrem{mobile} \revadd{not-fixed} solute \revadd{position}  can be regarded as a  homogeneous binary mixture, so that the following relation holds true
\be
\label{eq:muN}
 G_2 = \sum_{i=1}^2 \mu_i N_i = \mu_{solute} + \mu_2 N.
\ee
Up to the solute kinetic energy which anyhow cancels out in  free-energy differences, the value of $G_2$ in 
\revadd{eqs. \eqref{eq:G2} and \eqref{eq:muN} }
should be the same, 
so that using \cref{eq:Delta G fixed}
\be
\label{eq:mu solute}
 \mu_{solute} = \Delta \Theta_{MDFT} + {kT \over 2} n_0^2 \hat{c}(k=0) \Delta V.
\ee
Relations \eqref{eq:Delta G fixed} and \eqref{eq:mu solute}  constitute the main results of this paper. \revrem{Clearly,} \cref{eq:mu solute} \revrem{should be used when comparing to experiments}.
\revadd{
The \cref{eq:mu solute} was used for the MDFT calculations presented in this paper.}

All the above derivation stems naturally in a classical DFT context, in which the starting fundamental quantity is the free energy defined in the grand-canonical ensemble. As mentioned in the introduction, there have been much efforts in 3D-RISM approaches to correlate the deviation observed between the computed SFE and experimental or MD results to empirical partial molar volume corrections.  Since such integral equation approaches are developed in a grand-canonical framework by imposing the solvent density far from the solute, it can be proved under certain restrictions that the partial molar volume term derived above should apply {\em mutatis-mutandis} to the 3D-RISM calculations. This requires to extract the spherical component of direct correlation function, $\hat{c}_S(k)$, from a preliminary 1D-RISM calculation (see the supplementary material). 
Our DFT approach thus gives a sound theoretical justification to the empirical corrections that were proposed to date and  provides an  \revrem{unambiguous} \revadd{theoretical} value to the $\Delta V$-coefficient that should be used. In a very recent publication, Truchon et al\cite{truchon_cavity_2014} have proposed a so-called cavity correction term which looks formally at first sight very similar to ours in  \cref{eq:mu solute}. It is in fact quite different in nature: it involves the solute-solvent direct correlation function instead of the solvent-solvent in our formulation, and it requires an empirical multiplicative factor that is not present in our case. Moreover, our equation applies to a  mobile solute. It is not clear how the two approaches can be related.

\section{Results}

As an application of the MDFT formalism described above, and test of \cref{eq:mu solute}, we have computed 
the hydration free-energies of a series of 504 organic compounds, for which both experimental and molecular dynamics free-energy perturbation (MD-FEP) data are available from the work of Mobley et al.\cite{david_l_mobley_small_2009}. We  used as input the structures and partial atomic charges from the supporting information of this paper. This data set was used recently as a test case for interaction-site DFT\cite{liu_high-throughput_2013} and 3D-RISM \cite{truchon_cavity_2014} approaches.
 
Briefly, the procedure was as follows (with more details in the supplementary material). Each molecule was placed at the center of a cubic box with dimension 40 $\AA$, and MDFT minimizations were performed using the direct correlation function of TIP3P water\cite{TIP3P} (the model used in Ref.~\cite{david_l_mobley_small_2009}), and the  functional form and algorithms described in  Ref.~\cite{jeanmairet13}.
The position and orientation water density $\rhorom$ was represented on a 3D grid with  $100^3$ points for positions, and  an angular grid for orientations; we used a Lebedev grid of 6 orientations for the water molecular axis orientations (angles $\theta,\phi$), plus a 2-angles regular grid for the rotation around the molecular axis (angle $\psi$ from 0 to $\pi$). We used the functional and the minimization method described in Ref.~\cite{jeanmairet13}.
 Each minimization took about 10 minutes on a single CPU core.
8 out of the 504 molecules turned out to give divergent results and were discarded from the statistics. This failure is attributed to the shortcomings of the HRF approximation for molecules with very high local field. This could be corrected by three-body correction terms in the functional, that we do not consider in this paper.  All calculations below  were organized with the MolDB workflow system \cite{volodymyr_sergiievskyi_moldb:_2013}.
Statistical processing was performed with GNU Octave \cite{eaton_gnu_2005}.

\begin{figure}[h]
\begin{center}
  \includegraphics[width=0.8\textwidth]{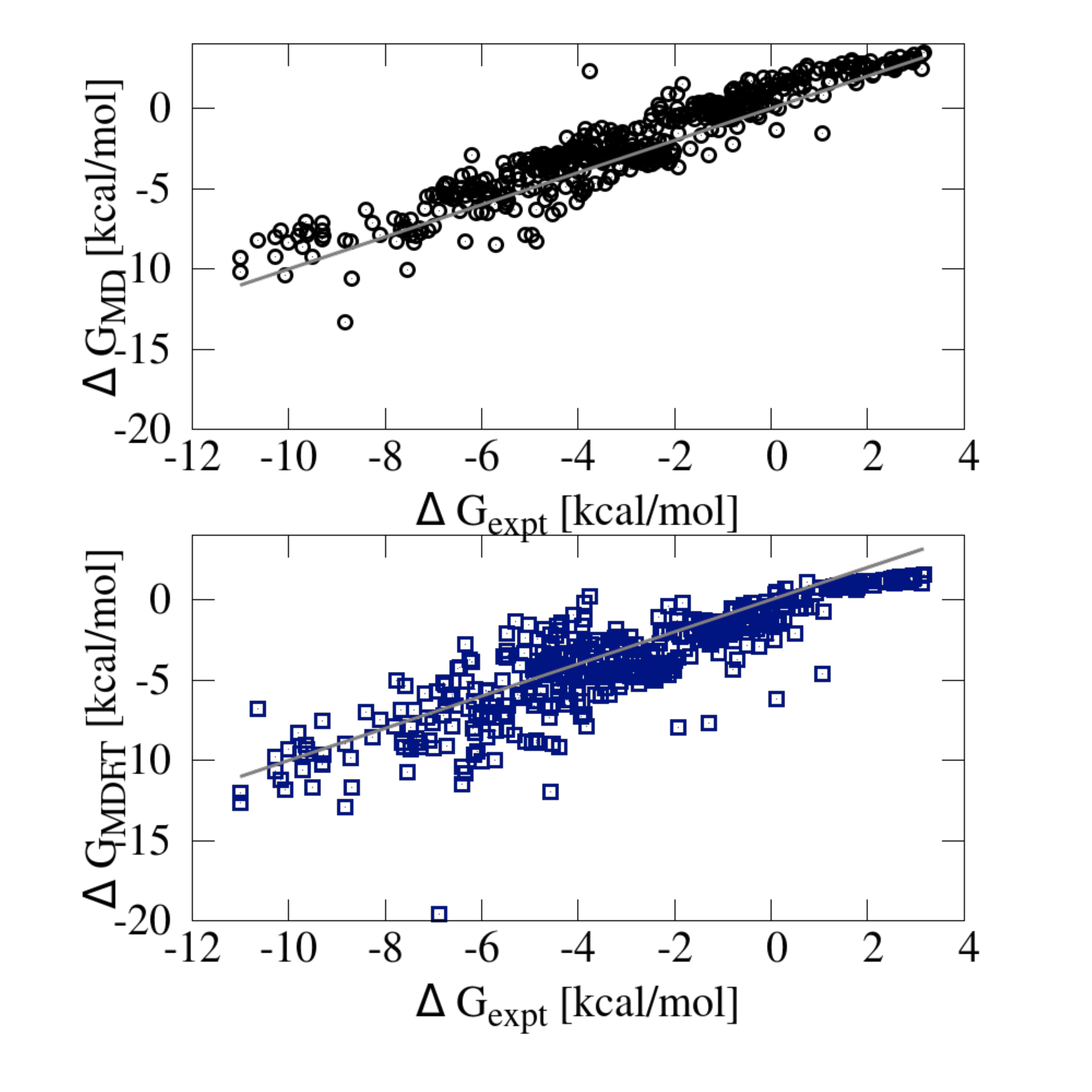}
\end{center}
  \caption{\label{fig:results_MDFT} MDFT results, converted to the $NPT$ ensemble (top) and MD results (bottom) correlated to  experimental results.
}
 \end{figure}

In Figure \ref{fig:results_MDFT} the results of the MD simulations and of the MDFT calculations are compared to the experimental solvation free energies.
We observe an evident correlation  of the MDFT calculations  with experimental data (correlation coefficient $r \simeq 0.9$).
The root mean square deviation (RMSD) of the MDFT results is 1.8 kcal/mol, which is  0.5 kcal/mol larger than that of the MD simulations (1.25 kcal/mol).
The higher dispersion of the results ($\sigma_{\text{MDFT}}$ = 1.6 kcal/mol instead of $\sigma_{\text{MD}}$  = 1.1 kcal/mol) points, again,  to  the shortcomings of the HRF approximation; See supporting information for more details.


\begin{figure}[h]
\begin{center}
  \includegraphics[width=0.8\textwidth]{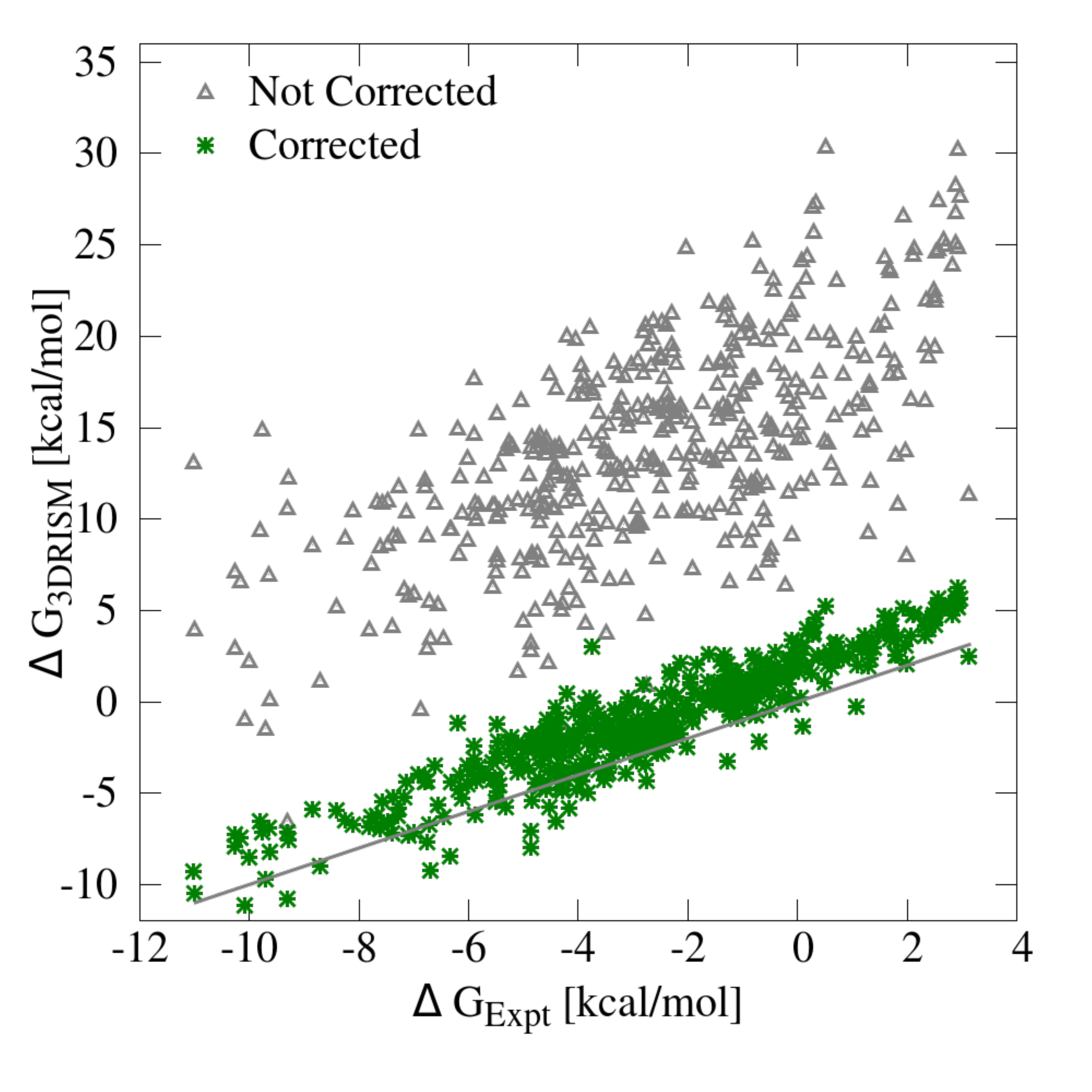}
\end{center}
\caption{\label{fig:results_3DRISM} 
3D-RISM solvation free energies with and without ensemble correction compared to  experimental results.}
\end{figure} 
 
For comparison, 3D-RISM calculations were also carried out for the same set of molecules. We used the multigrid 3D-RISM method of Sergiievsky et al\cite{sergiievskyi_3DRISM_2012}, that is available online \cite{volodymyr_sergiievskyi_rism-mol-3d:_2013}.
The calculations were performed on a rectangular grid with step size of \revrem{0.5} \revadd{0.2} \AA. The buffer (minimal distance from the solute to the boundary of the calculation box) was set to \revrem{12} \revadd{15} \AA.
\revadd{
The MSPC-E water model with additional LJ parameters of water hydrogen was used to describe the solvent in 3DRISM calculations. 
The following LJ parameters of water hydrogen were used: 
$\sigma_H$ = 1.1 \AA, $\epsilon_H$ = 0.046 kcal/mol.  
In the calculations the total site-site correlation functions of water calculated previously by Fedorov and Kornyshev which the dielectrically consistent RISM technique were used \cite{fedorov_unravelling_2007}.
To avoid divergence of the algorithm due to the long-range
behavior of the interaction potentials, we separate the short- and long-ranges of the potentials and then treat them separately by using the Ng procedure\cite{ng_hypernetted_1974}.
}
The HNC closure was used in the calculations. 
At the end of the calculation, we used \cref{eq:mu solute} to convert the results to the $NPT$ ensemble. The required function $c_S(r)$ was in this case expressed as a sum of site-site functions (consistent with the RISM assumptions).

The 3D-RISM calculations converged for \revrem{500}
\revadd{439} of 504 molecules \revrem{(convergence does occur for all molecules using the Kovalenko-Hirata (KH) instead of HNC closure)}
\revadd{The high divergence rate can be explained by the known pure divergence of HNC closure, and normally can be improved by using of the Kovalenko-Hirata (KH) closure.
However, in our opinion it is more consistent to use the HNC closure, as the theory was developed for the HRF approximation.
 }
These \revrem{500} \revadd{439} molecules were used for analysis.
The results are compared to the experimental data in Figure \ref{fig:results_3DRISM}.

We note first that the results without the ensemble correction of \cref{eq:mu solute} are very dispersed and poorly correlated to the experimental results. This is consistent with previous findings\cite{lue_liquid-state_1992,chuev_improved_2007}. It is also consistent with what is obtained with MDFT when  the ensemble correction is omitted; the rigorous DFT approach tells us however that this correction term has to be there. The RMSD of the 3D-RISM calculations appears slightly higher than that in MDFT (\revadd{2.39} kcal/mol).
We note, however, that most of this error is due to the systematic error ($M \simeq \revadd{1.9}$ kcal/mol), and that the standard deviation is close to that obtained with MDFT ($\sigma_{\text{3D-RISM}}$=\revadd{1.45} kcal/mol).
The relatively big systematic shift of the 3D-RISM results can be explained by the imperfect solvent properties used in the calculations (extracted from a preliminary RISM study of the bulk solvent)  and, obviously, by the shortcomings of the RISM approximations (see supporting information for details).

\section{Conclusions}

We have shown that the computation of solvation free energy with classical DFT, moreover in the framework of  the HRF approximation (equivalent to the HNC approximation  in integrals equation approaches),  should include a partial molar volume correction accounting for the change of ensemble from $\mu$VT to NPT. The proportionality factor is found to depend on $\hat{c}_S(0)$,  the value of the Fourier-transformed isotropic direct correlation function at $k=0$. This PMV correction was shown to be directly extendable to the 3D-RISM theory, with a  $\hat{c}_S(k)$ extracted from a consistent 1D-RISM description of the solvent. Although of related appearance,  it has a  different nature than the one proposed recently by  Truchon et al\cite{truchon_cavity_2014}.

Using this PMV correction, we have computed the solvation free-energies for the dataset of \revadd{504} organic molecules studied by Mobley et al.\cite{david_l_mobley_small_2009} using both MDFT and 3D-RISM with a HNC rather than KH closure. Since both methods rely on a related set of approximations,  they are found to yield an accuracy that is similar compared to experimental results, and only slightly worse than that obtained with much more costly MD-FEP simulations. Similar, \revrem{if not} \revadd{but} slightly better agreement were found for the same data set using interaction-site DFT\cite{liu_high-throughput_2013} or 3D-RISM with cavity corrections\cite{truchon_cavity_2014}; in both cases, a single parameter can be used to optimize the results (reference hard-sphere radius or multiplicative factor). 
\revadd{Although we accept the fact, that using one or several fitting coefficients the results could still be improved, we stress  importance of the parameter-free models (like in our work), because they are able to give a key to understanding of the underlying physical processes and thus can help to identify and improve week points of the model.}
\revrem{, whereas the present approach is completely parameter-free.
}

There is still much room and work to be done to improve the applicability and accuracy of the liquid-state theoretical methods. Certain improvements should be common to DFT and RISM, such as accounting for solute flexibility, which is naturally incorporated in MD-FEP methods, or accounting for solvent and solute polarisability. Other improvements should go separately in each approach. In integral equations, it is desirable to improve upon the HNC or KH closures that are limits of the theory nowadays. In DFT, it has been shown that the free-energy functional can be substantially improved by adding three-body correction terms,  either using a hard-core reference\cite{levesque12_1,jeanmairet13_2} or accounting for the non-additive character of H-bonding interactions\cite{zhao11}.  We have to determine how those further corrections will add to the  most basic one introduced in this paper: referring to the correct thermodynamic ensemble when comparing to experiment. 

{\bf Acknowledgements:} this work was supported by a grant from the Pierre-Gilles de Gennes foundation. The authors are very grateful to Jean-Fran\c{c}ois Truchon  for drawing their attention to the  data set by Mobley et al. (Ref.~\cite{david_l_mobley_small_2009}),  and kindly sharing a preprint of his latest work (Ref.~\cite{truchon_cavity_2014}) before publication. DB is also thankful to  Shuangliang Zhao for his wise contributions to the beginning and end of this work.

\end{document}